\documentclass[usenatbib]{mn2e}
\usepackage{times}
\usepackage{epsfig}
\usepackage{amsmath}
\usepackage{amssymb}
\title[Dust accretion and destruction in galaxy groups and clusters]{Dust accretion and destruction in galaxy groups and clusters}

\author[McGee et al.]{Sean L. McGee$^{1}$\thanks{Email:
    s2mcgee@uwaterloo.ca} and  Michael L. Balogh$^{1}$
\\
$^{1}$Department of Physics and Astronomy, University of Waterloo, Waterloo, Ontario, N2L 3G1, Canada\\
}

\date{\today}

\def\solarpc{$h \ {\rm M}_\odot / {\rm pc}^2$}
\def\LCDM{$\Lambda$CDM$~$}
\def\Mdoth{$h^{-1}~$M$_\odot$}
\def\Mpch{$h^{-1}~$Mpc$~$}
\def\kpch{$h^{-1}~$kpc$~$}
\newcommand{\kmsmpc}{\>{\rm km}\,{\rm s}^{-1}\,{\rm Mpc}^{-1}}
\begin{document}
\maketitle

\begin{abstract}
We examine the dust distribution around a sample of 70,000 low
redshift galaxy groups and clusters derived from the Sloan Digital Sky
Survey. By correlating spectroscopically identified background quasars
with the galaxy groups we obtain the relative colour excess due to
dust reddening. We present a significant detection of dust out to a
clustercentric distance of 30 Mpc/h in all four independent SDSS
colours, consistent with the expectations of weak lensing masses of
similar mass halos and excess galaxy counts. The wavelength dependence
of this colour excess is consistent with the expectations of a Milky
Way dust law with R$_V$=3.1. Further, we find that the halo mass
dependence of the dust content is much smaller than would be expected
by a simple scaling, implying that the dust-to-gas ratio of the most
massive clusters ($\sim$ 10$^{14}$ \Mdoth) is $\sim$ 3$\%$ of the
local ISM value, while in small groups ($\sim$ 10$^{12.7}$ \Mdoth) it
is $\sim$ 55$\%$ of the local ISM value. We also find that the dust
must have a covering fraction on the order of 10 $\%$ to explain the
observed color differences, which means the dust is not just confined
to the most massive galaxies. Comparing the dust profile with the
excess galaxy profile, we find that the implied dust-to-galaxy ratio
falls significantly towards the group or cluster center. This has a
significant halo mass dependence, such that the more massive groups
and clusters show a stronger reduction. This suggests that either dust
is destroyed by thermal sputtering of the dust grains by the hot,
dense gas or the intrinsic dust production is reduced in these
galaxies.

\end{abstract}


\section{Introduction}

Dust grains have long been known to play an important role in the
interstellar medium and the star formation which occurs within a
galaxy. Further, because dust grains can absorb and redden background
sources, an accurate knowledge of the large scale distribution of
these grains is crucial. Groups and clusters of galaxies present a
unique opportunity to study this large scale distribution as well as
the processes important in dust evolution. Dust within massive
clusters is thought to sputter on timescales of 10$^7$ - 10$^9$ years
\citep{draine_salpeter}, depending on the density and temperatures of
the environment. This is caused by the ejection of atoms from the dust
grain by the collision with sufficiently energetic gas
particles. Given the presence of a distributed, hot plasma in groups
and clusters, the short timescale for sputtering means that dust
observed in these systems must have been accreted recently. This
potentially gives a probe of the relevant dust creation processes.

Recently \citet{menard} have shown that dust excesses exist out to Mpc
scales around i$<$21 galaxies.  There are a variety of mechanisms by
which dust may escape from galaxies and become distributed on such
large scales.  Star-forming galaxies are often seen to have large
outflows of gas and dust which are blown out by the power from
supernova feedback \citep{heckman,tremonti_wind}. In addition, active
galactic nuclei (AGN) can have significant power, and the jets they
induce may be able to remove gas and dust from galaxies and
redistribute it within the larger environment
\citep[e.g.][]{mcnamara}. During galaxy-galaxy mergers a significant
amount of the gas and dust may be removed due to collisional
processes, or subsequently blown out by the induced star-formation or
AGN power \citep{hopkins}. Observations of galaxies falling into
massive clusters show gas and dust being stripped from the galaxy and
incorporated into the surrounding environments \citep{crowl,domainko}.

The temperatures and densities of the large scale environments of
groups and clusters are difficult to probe observationally. While
X-ray observatories have allowed the determination of the temperatures
and densities of massive clusters to near the virial radius
\citep[e.g.][]{vikhlinin}, analysis of the detailed properties of representative
samples of groups has largely relied on stacking large samples
\citep{dai} and/or been limited to measuring total X-ray luminosities
\citep{rykoff}. As a result, the densities and temperatures of the
environments outside the virial radius are largely inferred from
simulations alone \citep{pfrommer,kay}. Indeed, the non-detection of
significant amounts of baryonic material in the local universe has
given rise to the postulate that this material is contained in
overdense gas with temperatures of 10$^5$-10$^{7}$, dubbed the
warm-hot intergalactic medium (WHIM) \citep[see][]{bregman}. The
expected temperature and density of the WHIM are likely able to
destroy dust through thermal sputtering. Therefore, measuring the
large scale dust distribution is a potentially powerful probe of the
large scale temperatures and densities.

Previous attempts to measure the dust centered on clusters has focused
on the inner regions, largely within the virial radius. Observations
of typical groups ($< 10^{13.5}$ \Mdoth) have not yet been done. The
first attempts to quantify the dust content of clusters involved the
counting of relative deficiency of background sources.  These attempts
generally agreed that the $V-$band attenuation is on the order of
0.2-0.4 \citep{zwicky, bogart, boyle, romani}. However, early infrared
observations of clusters attempting to detect direct emission from
this dust largely led to non or marginal detections, implying dust
masses much lower than that implied by background counts
\citep{annis,wise,stickel,montier}. Similarly, recent observations
with Spitzer have also failed to show dust masses implied by 0.2-0.4
magnitudes of attenuation \citep{bai}.

Recently, large, uniform surveys have made detecting statistical
colour excesses of background sources of large samples of clusters
feasible. \citet{Chelouche} have correlated background quasars with
intervening galaxy clusters and found that, for massive clusters, the
excess reddening, E(B-V), is on the order of 0.004 magnitudes in the
central Mpc. \citet{bovy}, using the spectra of early type galaxies,
found an upper limit on the extinction within 2 Mpc of massive
clusters to be A$_V$ $<$ 0.003 magnitudes.  Similar results were
obtained by \citet{muller}, who found, using photometric redshifts to
identify background galaxies, a similar result of A$_V$ = 0.004 $\pm$
0.010 mag of attenuation. However, these type of measurements are
differential measurements and, as such, are dependent on the control
sample of sources with which the colour excess is measured.  Since it
is known that clusters of this halo mass have significant excess mass
out to at least 20 \Mpch \citep{sheldon1}, and indeed
\citet{Chelouche} has shown that the colour excess varies at least out
to 5 Mpc from the cluster center, then the presence of dust on these
larger scales may bias the measurement of the central regions.

In this paper, by measuring the reddening effect on background
quasars, we examine the large scale ($\sim$ 50 \Mpch) radial profile
of dust centered on groups and clusters. This is the first measurement
of such large scales and the first to probe such low mass groups. The
large scale is important to separate the cluster/group dust from
the dust expected to be associated with individual galaxies. We
discuss the data in \textsection \ref{data} and present the method and
the measurement in \textsection \ref{measurement}. We discuss the
implications of the measurement in \textsection \ref{discussion} and
summarize our findings in \textsection \ref{conclusions}. Throughout
this paper, we adopt, as was done during the assembly of the group
catalogue, a \LCDM cosmology with the parameters of the third year
WMAP data, namely $\Omega_{\rm m} = 0.238$, $\Omega_{\Lambda}=0.762$,
$\Omega_{\rm b}=0.042$, $n=0.951$, $h=H_0/(100 \kmsmpc)=0.73$ and
$\sigma_8=0.75$ \citep{spergel}.

\section{Data}\label{data}

The technique we use to examine the dust content of galaxy groups and
clusters is conceptually simple.  Our approach relies on measuring the
change in the mean colour of background sources as a function of their
projected distance from groups and clusters. It is expected that the
change in the mean colour will be small, and thus we must stack the
signal from many clusters together to measure the effect on the
distribution of background sources. Therefore, we require a large
sample of groups and clusters, which are uniformly selected and have
well defined masses. We also require that we have a large number of
background sources with an intrinsic colour distribution with little
scatter and which do not vary with sky position. The background
sources must also be at high redshift, so that they are not physically
associated with the clusters we are examining.  For these purposes,
the best publicly available data is derived from the the Fourth Data
Release (DR4) of the Sloan Digital Sky Survey, a five colour ($ugriz$)
photometric and spectroscopic survey covering over 4780 deg$^2$ and
containing $\sim$ 670,000 spectra of galaxies, quasars and stars
\citep{sdssdr4_short}. Below we discuss the group sample and the
sample of background objects drawn from this survey which are used in
this paper.

\subsection{Galaxy group catalogue}

The majority of galaxies in the local universe reside in some kind of
association with at least one other galaxy \citep{eke_2pigg}. However,
quantifying the mass of a large sample of those associations is
difficult. In particular, X-ray temperature or luminosity, often used
as a mass indicator in galaxy clusters \citep{reiprich}, is too low or
faint in galaxy groups to determine mass for a survey the size of
the SDSS. Also, the velocity dispersions of galaxies, which are often used
as tracers of the potential well of clusters or massive groups, is
ineffective when the groups contain only a handful of
spectroscopically confirmed galaxies. Recently, \citet{YMvJ} showed
that an effective mass estimate of galaxy groups can be obtained by
essentially ranking groups by their total luminosity and associating
these rankings with the expectations of a \LCDM halo occupation model.

\begin{figure}
\includegraphics[width=8cm]{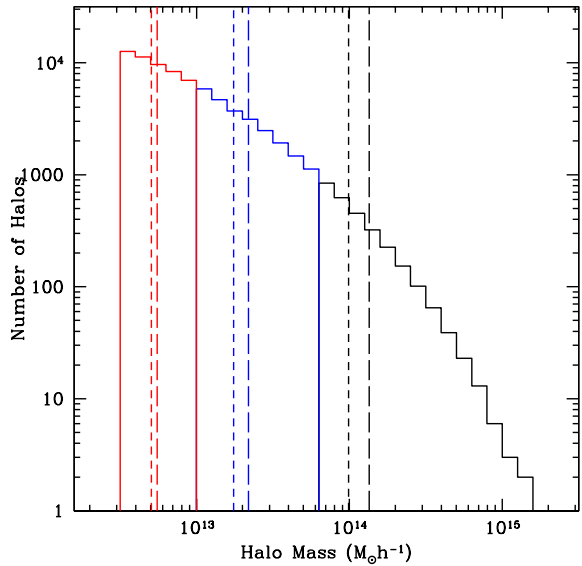}
\caption{The halo mass distribution of the galaxy group and cluster
 sample. The sample is shown in three groups of halo mass, from
 10$^{12.5}$ to 10$^{13}$ \Mdoth (red),  10$^{13}$ to 10$^{13.8}$
 \Mdoth (blue), and 10$^{13.8}$ to 10$^{15.3}$ \Mdoth (black). The
 short dashed (long dashed) line in each bin represents the median
 (mean) halo mass of the bin.
 }
\label{halomassgroups}
\end{figure}

\citet{yanggroups} have applied their algorithm to the SDSS DR4 to
produce a sample of $\sim$ 300,000 galaxy groups with masses as low as
10$^{11.5}$ \Mdoth. In this paper we use ``Sample I'' from
\citet{yanggroups}, which exclusively uses galaxies with SDSS
spectroscopic redshifts. We chose this sample, as opposed to the other
samples, which add in existing redshifts from the literature, because
we are principally concerned with obtaining a uniformly selected
population of galaxy groups. In the most recent galaxy group catalogue
of \citeauthor{yanggroups}, the authors rank the galaxy groups both by
total luminosity in the $r-$band as well as the total stellar mass of
the group. In this paper, we will use the group masses obtained by the
ranking the total stellar mass, with the goal of minimizing the effect
of a particular group's recent star formation history. However, we
note that because we must stack many groups in relatively wide mass
bins, this choice has no effect on the results. Finally, we restrict
our analysis to galaxy groups with masses greater than 10$^{12.5}$
\Mdoth, which leaves a final sample of 75947 groups between z=0 and
z=0.2. Figure \ref{halomassgroups} shows the halo mass distribution of
our group sample broken into three mass bins which we use later in the
paper.

\subsection{Background objects: spectroscopically identified quasars}

Our principal concerns when choosing a sample of
background objects to examine are that the sources are at sufficiently
high redshift that they are not physically associated with the group,
that they represent a relatively uniform population, with a small
dispersion about their average properties, and that their colours are
well calibrated. For all these reasons we have chosen to examine the
photometric properties of spectroscopically identified quasars drawn
from the fourth edition of the SDSS quasar catalog
\citep{dr5_qso}. This catalog contains $\sim$ 77,000 quasars brighter
than M$_i=-22$ drawn from the fifth SDSS data release, and are
targeted based on their photometric properties and/or the presence of
an unresolved radio source \citep{sdss_quasar}. We reduce this sample
to contain only quasars within DR4, ie. those for which we have an
identified group sample. The SDSS quasar catalog contains {\it psf}
magnitudes for each of the five bands of SDSS photometry (${\it
ugriz}$), with typical errors of 0.03 mag. The colours have been
corrected for galactic extinction using the dust maps of
\citet{dustmaps}. We only examine quasars with redshifts between z = 1
and z = 3 in order to avoid quasars physically associated with the
target groups, or extreme objects at high redshift.

\begin{figure*}
\includegraphics[width=\textwidth]{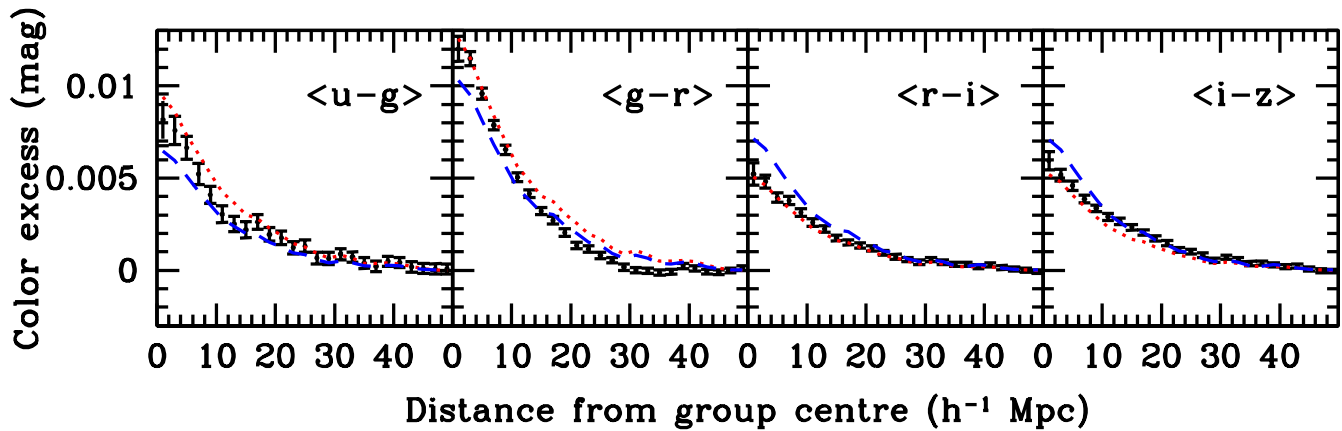}
\caption{The excess colour of four independent colours as a function
  of distance from the group center. The excess colour is measured
  with respect to the colour of quasars at a projected distance of 46
  to 50 \Mpch. The error bars are 1 sigma errors from Monte Carlo
  estimation. The blue, dashed line (red, dotted line) represents the
  R$_V$=2.0(5.0) dust reddening law which minimizes the chi-square for
  A$_V$ as a function of radius.  }
\label{color_excess}
\end{figure*}

\section{Analysis}\label{measurement}

Our first goal is to measure the change in the mean colour of the
background quasars as a function of their position from foreground
clusters.  This change in the mean colour is also known as the excess
reddening because interstellar dust is observed to extinguish light
more readily at the blue end of visible light, causing the light to
redden.

We parametrize the excess reddening, $E(i-j)$, of a source, $k$, as 
\begin{equation}
E( i - j )^k  =  m_i^k - m_j^k  - \langle m_i -
m_j \rangle_{control}
\end{equation}
where $m_i$ and $m_j$ are the magnitudes in the $i$ and $j$ bands of
the source for which the excess reddening is measured. The final term
in this expression is the mean colour of the control sample of
objects. In practise, the excess reddening measured in this way of a
single background source is dominated by the intrinsic distribution
width of colours of that source. However, with the assumption that the
background sources have a well defined mean colour we can stack large
numbers of sources to measure a mean excess colour, $\langle E(i-j)
\rangle$. Therefore, by stacking many backgrounds sources in bins of
projected distance, $r$, from the centre of our group and cluster
samples, we can obtain an excess colour profile of the groups and
clusters, $\langle E(i-j) \rangle (r)$.

A key feature of this type of measurement is the control sample of
sources. Unless the unobscured mean colour of the background sources
is known, it must be made as a differential measurement with respect
to a control sample of background sources. In the literature, the
control sample is not well defined. \citet{bovy} use a control sample
of all background sources greater than 2 \Mpch from the center of a
cluster, while \citet{Chelouche} defines the control sample to be
background sources which are greater than 7$R_{200}$ from the cluster
center. We have, by trial and error, found that the differential dust
signal approaches zero only at $>$40 \Mpch from the center of the
cluster. So, the control sample of background sources is must be taken
from large radius.  The control sample of quasars are defined as those
which have projected clustercentric distances, $r$, of 46 \Mpch $<$
$r$ $<$ 50 \Mpch from a given cluster in the sample.

Notice that, similar to the weak lensing or cluster correlation
measures, our dust measurement is a correlation measurement. A given
quasar can have a small projected distance from cluster A, but still
be counted as a control quasar if it is in the control range of cluster
B. In other words, each quasar is in many clustercentric radial bins.
Because of the much larger area covered by the control sample of
quasars, the number of quasars in the control sample greatly outnumber
the number of quasars in any of the other samples. This enables us to
do a Monte Carlo estimation of the error in the excess colour by
selecting 200 independent samples of control quasars which have the
same number as a given target bin. The error bars are then the 1 sigma
limits of the excess color from the 200 trials for each bin.  

\subsection{Large scale distribution}

We begin by examining the total colour excess from a stacked set of
all the groups in our sample. We measure the mean colour of the
background quasars within clustercentric annuli in projected radius
bins of 2 \Mpch, out to 50 \Mpch from the cluster center. The five
band photometry of the SDSS quasar sample allows for the measurement
of four independent excess reddening signals. Figure
\ref{color_excess} shows the colour excess for each of the four sets
of independent colours.  

As Figure \ref{color_excess} shows, a significant colour excess is
measured in all four independent colours as far out as 30 \Mpch from
the group and cluster centre. The typical projected virial radius of a
group or cluster of this mass (M $\sim$ 10$^{13.5}$ \Mdoth) is only
$\sim$ 1 \Mpch. Thus, such a significant large scale distribution of
reddening is initially surprising. However, studies of the cluster
correlation function \citep{bahcall}, and the weak lensing profile of
similar clusters \citep{sheldon1, johnston} shows that there is a
significant excess of matter out to similar distances from
clusters. We expect that the dust excess to such large scales is just
a result of this excess matter, although we will address this further
in \textsection \ref{dustgal}.

While the size of the measured excess reddening shown in Figure
\ref{color_excess} is quite small ($<$ 0.015 mag), if it is due to the
presence of dust then it corresponds to a large and extended dust
distribution. However, there are some systematic effects which 
must be accounted for before we can be sure the reddening is due to
dust. In particular, galaxy groups and clusters have significant mass,
and therefore sources behind this mass will be gravitationally
magnified. A sample of sources which is chosen by a fixed flux limit
will lead to more high redshift sources, and a lower absolute
luminosity limit at fixed redshift behind the cluster, than for a
patch of sky far from the cluster. Thus this measurement is
potentially affected when the intrinsic mean colour of the background
sources is luminosity or redshift dependent.

We can estimate the size of the magnification effect of our groups and
clusters. We would expect this effect to trace the mass distribution,
so it would be most pronounced close to the cluster. \citet{johnston}
measured the mass profile of a large sample of galaxy clusters and
found that at a distance of 1 \Mpch from the most massive clusters the
surface mass density is $\sim$ 10$^{2}$ \solarpc. This corresponds to
a magnification of $\sim$ 0.02 magnitudes --- a minute change in the
effective limiting magnitude of our quasar sample. Given this, we see
that for our $<$g-r$>$ measurement of $\sim$ 0.011 at 1 \Mpch to be
completely explained by magnification of a quasar sample with varying
mean colour, the intrinsic ${\it cumulative}$ $g-r$ colour of our
quasar sample would have to vary by 0.55 per magnitude near the
magnitude limit. However, we find that the cumulative $g-r$ colour of
the control sample of quasars only varies by 0.005 magnitudes from
i=19 to i=20.2, the magnitude limit of our sample. In other words, as
it is for all four independent colours, the size of the
magnification-induced reddening effect is two orders of magnitude
smaller than required to explain our results.

We note that the possibility of having foreground emission from the
group or cluster contribute to the the reddening signal is not
physical, given that the local background subtraction used in the
quasar photometry will remove this component.

\subsection{Wavelength dependence of reddening}

Here we consider the wavelength dependence of the reddening, to gain
some indication of the nature of the dust. The wavelength dependence,
$R_V$ is commonly parametrized by linking the absolute extinction in
the $V$ band, $A_V$ to the excess $B-V$ colour as
\begin{equation}
R_V=\frac{A_V}{ E(B-V)}.
\end{equation}
A value of $R_V = 3.1$ is generally used based on studies of
interstellar Milky Way dust, but values in the range from 2.5 to 5
have been measured for the Magellanic clouds \citep{prevot} and
starbursting galaxies \citep{calzetti}.  In contrast, the emission
from a typical collection of cluster galaxies would have a negative
R$_V$. Unfortunately, we can not directly determine $R_V$ because we
do not have $B$ and $V$ magnitudes. However, using the fitting
functions of \citet{odonnell}, we can transform our excess colours to
determine $A_V$ and to determine whether the shape of the expected
reddening curve is consistent with the known properties of
interstellar dust.

In Figure \ref{color_excess}, we also show the predictions of two dust
reddening laws, one with R$_V$ =2.0 and one with R$_V$=5.0. To do this
we must find the A$_V$ value at each radial bin from the combination of
the four independent colours. We do this by minimizing the chi square
value at each step given the errors on each colour. For the majority
of the colours and radii, these two dust laws essentially bracket the
data. This seems to suggest that the colour profile is very similar to
that expected from known dust laws. However, we would like to make a
more precise measurement, which we do by stacking the data in radial
bins.

\begin{figure}
\includegraphics[width=8cm]{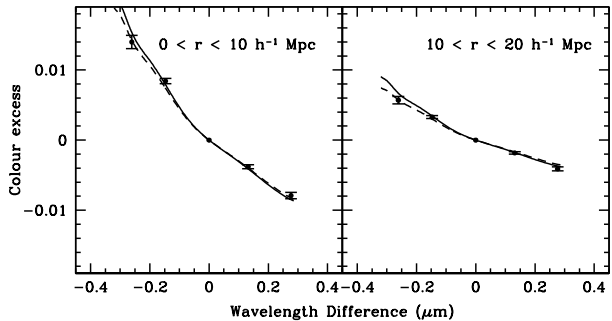}
\caption{The colour excess in two bins of radial distance from the
  group centres. The solid black line is the expectations from a
  R$_V$=3.1 dust law, using the expansion of \citep{odonnell}, with a
  attenuation of A$_V$ = 0.0223 and 0.0097 for the 0 $>$ r $>$ 10
  \Mpch and 10 $>$ r $>$ 20 \Mpch bins respectively. The dotted line
  is the result of simultaneously fitting for R$_V$ and A$_V$, which
  results in (R$_V$,A$_V$) = (3.3, 0.0220) and (3.5, 0.0094) in the
  two radial bins, respectively. The data points are for the colours
  u-r, g-r, r-r, i-r, z-r respectively.}
\label{dust_law}
\end{figure}

In Figure \ref{dust_law}, we show the reddening of background
quasars in two radial bins for each photometric band with respect to
the r band. The solid black line in each panel is the extinction curve
expected from an interstellar dust law with R$_V$=3.1. The dust
attenuation, A$_V$, is 0.0223 in the panel showing the inner 10 \Mpch
, while it is 0.0097 for the 10 $>$ r $>$ 20 \Mpch bin. While the
R$_V$=3.1 law shows reasonable agreement with the wavelength
dependence of the reddening, we also allow R$_V$ to be fit
simultaneously with A$_V$. These results are shown with the dotted
line and correspond to best fitting R$_V$ values of 3.3 and 3.5 in the
two radial bins respectively. Thus, the reddening is very similar to
the expectations of a Milky Way dust law.

\subsection{Halo mass dependence}
\begin{figure}
\includegraphics[width=9cm]{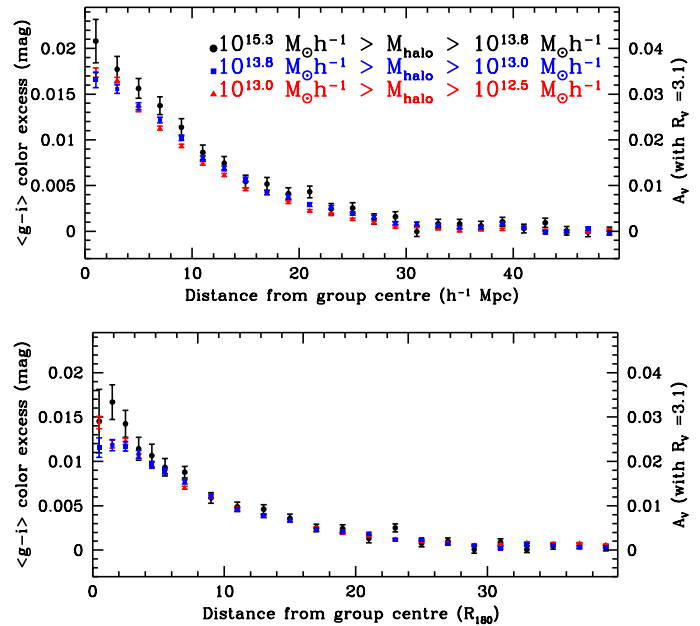}
\caption{The $g-i$ colour excess in three samples of galaxy groups as
  a function of distance from the group center. The V band attenuation
  $A_V$ is plotted on the right ordinate axis assuming R$_V$=3.1. This
  is shown as a function of physical distance (top panel) and in
  R$_{180}$ (bottom panel).
  }
\label{gi_color_excess}
\end{figure}

We now explore how this dust distribution depends on the mass of the
host halo.  In Figure \ref{gi_color_excess} we show the qso $<$g-i$>$
colour excess as a function of distance from the group center in three
bins of total group mass. The $<$g-i$>$ colour is chosen as our
principle measure of the dust reddening in the remainder of the
paper. These two photometric bands are the best calibrated, and have
been recently used in the literature. Conveniently, the $<$g-i$>$ is
most closely related to the more commonly used $B-V$ dust colour as
\citep{prevot, menard_mg}

\begin{eqnarray}
E( g - i ) &=& \frac{\lambda^{-1.2}_g -
  \lambda^{-1.2}_i}{\lambda^{-1.2}_B - \lambda^{-1.2}_V} E( B - V )
\\
&=& 1.55  E( B - V ).
\end{eqnarray}

Using this transformation, and assuming $R_V$=3.1, we show the
corresponding dust attenuation in the $V$ band, $A_V$, on the right
ordinate axis. Quite strikingly we see that the centers of the
clusters have dust attenuations of $A_V$ $\sim$ 0.03-0.04, when
measured with respect to quasars $\sim$ 45 \Mpch away. In comparison,
the disk of a spiral galaxy at a similar redshift has a dust
attenuation of $A_V$ $\sim$ 0.25 mag \citep{holwerda}. This suggests
that the dust signal cannot be localized to disks within the line of
sight alone, because the covering fraction of such group and cluster
members is much less than $(0.03/0.25)=0.12$. In other words, there
must be a significant component of dust which is not localized within
the disks of $\sim$ 0.1L* galaxies alone. A similar conclusion was
reached recently by \citet{menard}, who showed that the dust
attenuation around galaxies extends well beyond the radius expected if
the dust was confined to the disk. We examine this further in
\textsection \ref{sec-spatial}.

It is also noticeable that the dust distribution shows a relatively
small halo-mass dependence, in that the most massive bin shows a
larger colour excess than the lowest mass bin at each position out to
20 \Mpch. However, except at the inner most bins, the measured colour
excess in each of the three mass bins is always within $\sim$
0.001--0.002 magnitudes, a much smaller range than the range of their
values as a function of group centric distance.  A possible
explanation for such a small dependence on the halo mass could be that
the halo masses themselves are uncertain, so that perhaps the wide
halo mass bins actually contain roughly the same size groups. However,
it is worth noting that the fraction of red galaxies within these same
groups show a significant halo mass dependence. 50$\%$ of the $L_*$
galaxies in 10$^{14}$ \Mdoth are ``early type'' while only 20$\%$ of
the same galaxies are ``early type'' in 10$^{12.5}$ \Mdoth haloes
\citep{weinmann}. This argues that the masses are well defined for a
statistical study such as this.  

In the bottom panel of Figure \ref{gi_color_excess}, we show the same
$<$g-i$>$ colour excess but now as a function of scaled radius, namely
R$_{180}$. For this plot we have used the individual R$_{180}$ of each
cluster, rather than using the median R$_{180}$ of each bin. This is
an important point, as simply scaling physical distance in the top
panel by the median R$_{180}$ would result in lower mass groups
actually have a higher reddening than higher mass groups. However,
because there is a range of R$_{180}$ in each bin, the result
is to almost completely remove the halo mass dependence, especially at
distances far from the cluster. It appears that the dust reddening is
similar at a given scaled radius far from the cluster. We will return
to the interesting behavior at small scales in \textsection \ref{dustgal}.

\subsection{Spatial distribution of the dust}\label{sec-spatial}

By measuring the quasar colour excess as the difference in the mean
value of the quasars, our results are insensitive to the spatial
distribution of the dust. If the dust exists in concentrated
clumps with a small covering fraction, then the colour
excess could be driven by relatively few objects with large reddening
values. However, if the dust is uniformly distributed we would expect
each quasar to be reddened by the mean value. 

In an attempt to address this question, we explore the color excess as
a function of the percentiles of the quasar color
distributions. In order to determine what fraction of quasars
contribute to the signal, we sort the quasars in a
particular radial bin and those in the ``control'' background sample
by their measured $g$-$i$ colours. We bin each sample by
percentile and measure the mean $g$-$i$ colour in that percentile
bin, and then subtract the measured mean $g$-$i$ percentile colour of
the cluster quasars from the measure mean $g$-$i$ colour of the
``control'' sample for in the corresponding percentile. This leaves us
with the $\langle g - i  \rangle$ colour excess in percentile bins.

\begin{figure}
\includegraphics[width=9cm]{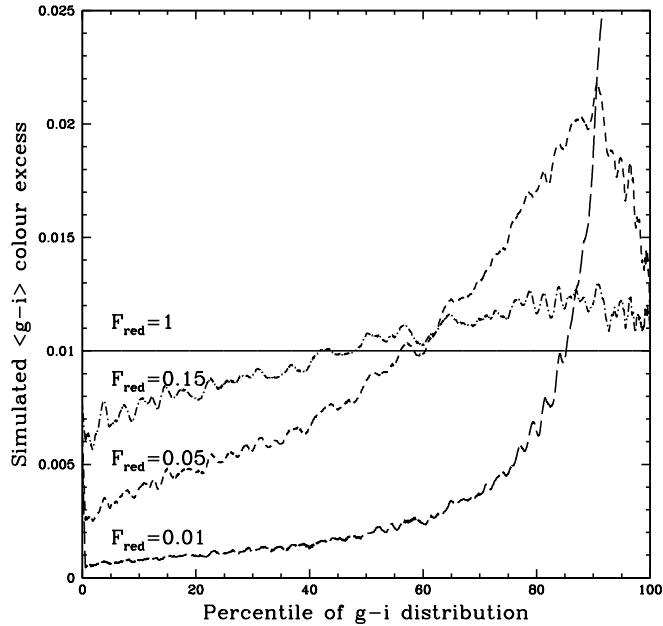}
\caption{The simulated $g-i$ colour excess as a function of percentile
for four different reddening models. The model is set to have a mean
color excess of 0.01 and F$_{red}$ determines the fraction of objects
which are reddened. }
\label{sim_cover}
\end{figure}

Admittedly, this measure is difficult to interpret physically, so we
first attempt to develop a framework in which to interpret the results
by using simulated reddening on background quasars. We take as our
control sample of quasars those which are the control sample of the
10$^{13.5}$-10$^{13}$ mass bin. We then add an amount of reddening to
some fraction of the background quasars (F$_{red}$) such that the mean
colour excess is 0.01; for instance, all quasars can be reddened by
0.01 magnitudes, or 1/3 of the quasars can be reddened by 0.03
magnitudes. The results are shown in Figure \ref{sim_cover}, where
F$_{red}$ is varied to be 1, 0.15, 0.05 and 0.01. As expected, the
color excess as a function of color percentile is flat for F$_{red}$=1
and becomes more dominated by a high percentile peak with lowered
F$_{red}$.

\begin{figure*} 
\includegraphics[width=\textwidth]{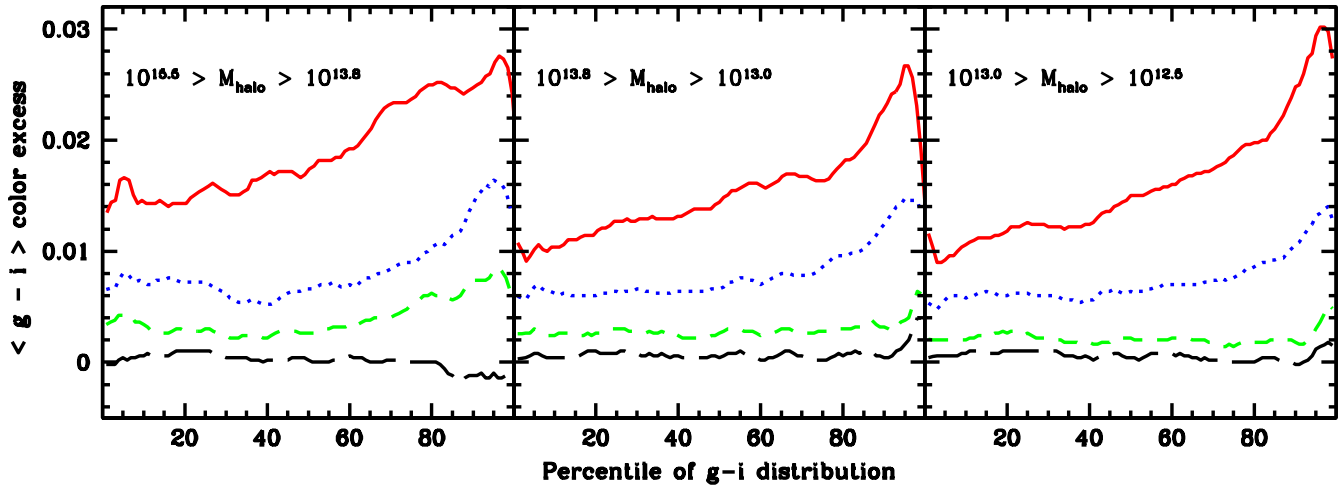}
\caption{The relative color excess at a given percentile of the quasar
  color distribution. This is shown in several radial bins for each
  halo mass range. The color excess is measured with respect to the
  color at a given percentile of the control quasars. The red, solid
  line is for quasars within the central 2 \Mpch, while the blue,
  dotted line is for quasars within  9 $<$ r (\Mpch) $<$ 11. The
  green, dashed (black, long dashed) line is for quasars within 19 $<$
  r (\Mpch) $<$ 21 (29  $<$ r (\Mpch) $<$ 31).}
\label{dust_cover}
\end{figure*}

We now present Figure \ref{dust_cover}, which shows
the color excess as a function of the percentiles of the real quasar color
distributions. This is shown for several radial bins and for each halo
mass range. We notice that for the majority of the radial bins the quasar color
excess is relatively constant, never more than $\sim$ 50 $\%$ away from the
mean value in that bin. Comparing these with the results of our
simulated data suggest that $\sim$ 15 $\%$ of the quasars are causing
this signal. Further, we see that the color excess curves become
progressively steeper with decreasing radius, implying that the bulk of
the signal is coming from fewer objects as we get near to the
cluster. In the inner-most bin the reddening can be explained with
5--15\%  of the quasars dominating the signal. 

As we will see in section \textsection \ref{exsection}, there are
$\sim$ 20 galaxies ($i$ $<$ 21) in the central 2 \Mpch of our most
massive clusters. Thus, assuming that the reddening signal is
concentrated uniformly in ``bubbles'' around these galaxies, a 10$\%$
covering fraction requires the ``bubbles'' to have radii of $\sim$ 140
\kpch. These bubbles are significantly larger than the typical size of
the stellar mass in a galaxy \citep{shen_size, mcgee} . So, while this
excess dust is not isotropically distributed throughout the groups, it
also is not confined to massive galaxies.

The results of Figure \ref{dust_cover} are also relevant in
understanding a possible selection effect on our results. Our colour
excess measurement are based on samples from a magnitude limited
survey, and thus dust extinction may move quasars out of our sample,
thereby leaving a relatively reduced color excess. However, examining
Figure \ref{dust_cover} we see that there does not exist any
significant population of highly extincted quasars. Further, this
result exists even when the sample is reduced to different magnitude
bins. The lack of a highly extincted population suggests that this is
not a significant effect for the bulk properties of the sample.

\section{Discussion}\label{discussion}

We have shown that there is a significant distribution of dust
centered on galaxy groups and clusters and, in this section, we would
like to measure some of its physical properties. This requires more
information about the physical nature of the dust and, while we were
able to show that the reddening is consistent with that expected from
local interstellar dust, our measurements are somewhat crude. In what
follows we will therefore make the assumption that the dust is similar
to Milky Way interstellar dust with an $R_V$=3.1.

\begin{figure}
\leavevmode \epsfysize=8cm \epsfbox{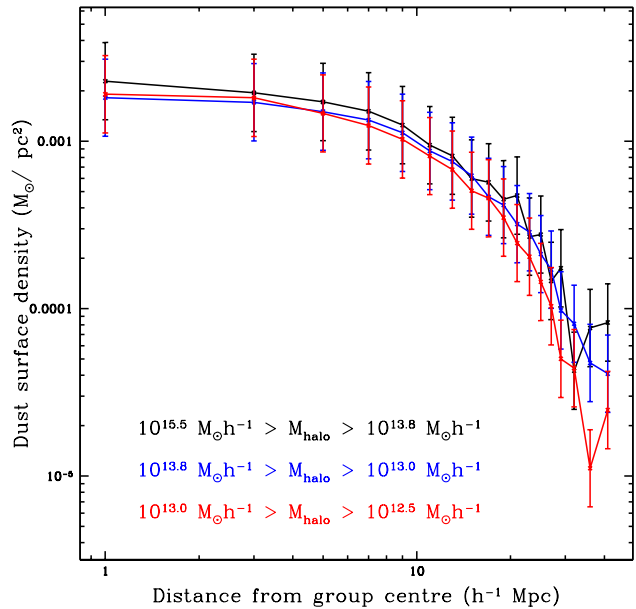}
\caption{The excess dust surface mass density, in units of \solarpc ,
 as a function of group-centric distance. This is shown for three
 different bins of group mass.  }
\label{surdens}
\end{figure}

We are most interested in converting the $A_V$ values we have measured
into a dust mass surface density. To do this we need an estimate of
K$_{ext}$, the extinction cross section for a given mass of dust:
\begin{equation}
\Sigma M_{dust} = \frac{A_V}{K_{ext}}
\end{equation}
For this, we use the carbonaceous-silicate dust grain model developed
by \citet{weingartner} and \citet{li_draine} and subsequently tweaked
by \citet{draine}. These models have been shown to successfully
reproduced the observed infrared emission, scattering properties and
extinction properties of local interstellar dust. The value of
$K_{ext}$ is wavelength dependent, but in the $V$ band it is
$K_{ext}=1.54 \times 10^4 cm^2/g$. With this assumption we can now
directly show the surface mass density of dust which is responsible
for the reddening signal. We plot this in Figure \ref{surdens} for
each of the three halo mass bins.

From this we conclude that the dust mass located within 1 \Mpch of
the cluster center is on the order of 10$^{10}$ \Mdoth, with little
dependence on halo mass.  If we assume that the gas fractions of these
clusters are equal to the universal baryon fraction this implies that,
for a 10$^{14}$ \Mdoth\ cluster, the dust-to-gas ratio is $\sim$
0.0003, or 3$\%$ of the interstellar medium ratio
\citep{Pei}. However, for a 10$^{12.5}$ \Mdoth group, the ratio is
$\sim$ 0.0055, approximately 50$\%$ of the interstellar medium ratio.

Here it is worth examining the theoretical expectations of thermal
sputtering. \citet{draine_salpeter} have shown that, for dust grains
principally composed of graphite, silicate or iron, collisions
with gas of 10$^6$ $<$ T $<$ 10$^9$ K can destroy them. Then, for gas
of this temperature dust grains can have a typical lifetime $\tau$ of 
 
\begin{equation}
\tau \approx 2 \times 10^4 \rm{yr}\left(\frac{cm^{-3}}{n_H}\right)\left(\frac{a}{0.01 \mu
  m}\right)
\end{equation}
where $n_H$ is the gas density and $a$ is the radius of the typical
dust grain. For the more massive clusters, a typical gas density is a
few 10$^{-3}$ cm$^{-3}$ while interstellar grains are smaller than 0.5
$\mu m$ \citep{mathis}. Therefore, dust grains in this gas will only
survive for $\sim$ 1 Gyr. So any dust which exists in the more massive
clusters must have been accreted within the last
Gyr. \citet{mcgee_accretion} have shown, using semi-analytic merger
trees, that local galaxy clusters accrete $\sim$ 5$\%$ of their final
galaxies per Gyr from otherwise isolated halos. This suggests that
only 5$\%$ of the total accreted dust should exist in massive clusters
today. This is remarkably similar to the implied dust-to-gas ratio of
the most massive clusters.

\subsection{Excess galaxy profile}\label{exsection}

\citet{menard} have recently shown that the dust-to-total mass ratio
is approximately constant to a distance of 10 Mpc from individual
galaxies. Similarly, they have shown that the dust-to-galaxy ratio is
constant, except within the inner 20 kpc of the galaxy, where it's
expected the signal is due to the galactic disk. We would like to
explore similar ratios as a function of group centric
radii. Unfortunately, we do not have available weak lensing mass
profiles for our groups and clusters. However, we are able to
correlate the excess number of galaxies in the SDSS photometry
catalogues with the positions of our groups and
clusters. Additionally, it is worth pointing out that \citet{sheldon}
has found that, for clusters similar to ours, the weak lensing mass to
light ratio is approximately constant from 1 Mpc to $\sim$ 20
Mpc. Therefore, we would expect that the galaxy/dust ratio should show
similar trends to the mass/dust ratios.

\begin{figure}
\leavevmode \epsfysize=8cm \epsfbox{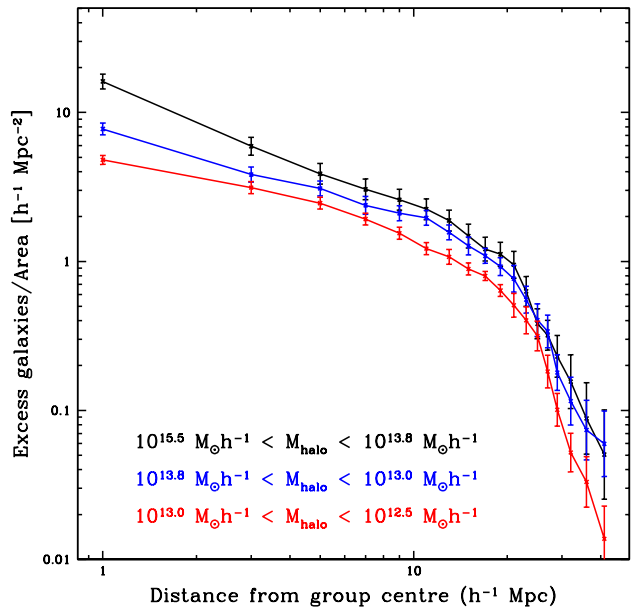}
\caption{The excess number of galaxies  ($i < 21$) as a function of
 cluster-centric position for three bins of halo mass. The excess
 number was determined by statistical background subtraction. The
 uncertainties are 1 sigma expectations from the Jackknife method.
 }
\label{excess_gals}
\end{figure}

In Figure \ref{excess_gals}, we present the excess number of galaxies
as a function of cluster-centric position for each of the three bins
of halo mass. This measurement was made by correlating the position of
all SDSS galaxies with a galactic de-reddened $i$ magnitude of $<$ 21,
where special care was taken with survey edges by using the SDSSpix
code.\footnote{http://dls.physics.ucdavis.edu/$\sim$scranton/SDSSPix/} We
choose to use the magnitude limit of $i$ $<$ 21 because this limit was
used by \citet{menard}.  This figure demonstrates that the excess
number of galaxies has a steeper slope in the inner 10 \Mpch than the
similar excess dust mass plot. The figure also shows that there are
significantly more galaxies in the higher halo mass bins, which again
argues that the halo masses are well defined for large statistical
samples.

\subsection{Dust mass associated with each galaxy}\label{dustgal}

\begin{figure}
\leavevmode \epsfysize=8cm \epsfbox{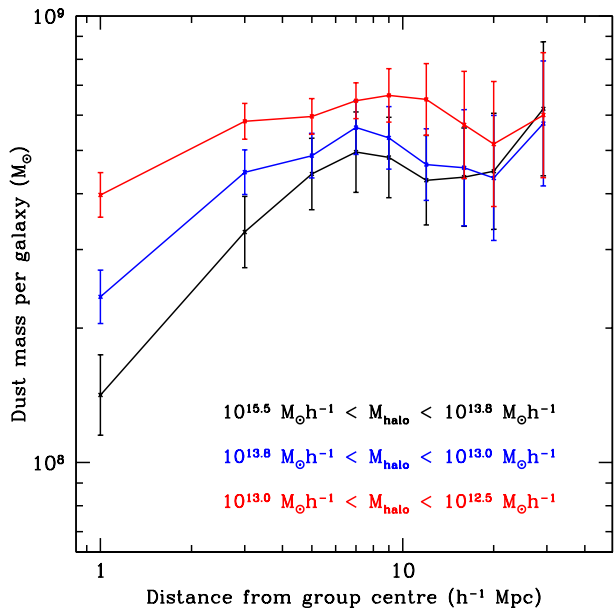}
\caption{The excess surface dust mass per excess galaxy as a function
 of distance from the group centre. 
 }
\label{masspergal}
\end{figure}

Now that we have compiled the excess dust mass distribution around the
clusters and the excess galaxy number around the clusters, we can
examine their ratio. We show this in Figure \ref{masspergal}. There
are a number of striking aspects of this plot. In particular, we see
that for the bulk of the galaxies, those which are $>$ 7 \Mpch away,
the excess dust mass per galaxy is on the order of 10$^{9}$
M$_\odot$.  This is approximately equal to the stellar mass
of the median excess galaxy, and roughly agrees with supernova
feedback models, which imply that up to 50$\%$ of metals produced are
blown out of the galaxy \citep{finlator}. The constant dust per
  galaxy fraction also confirms our earlier suggestion that the dust
  excess seen out to large radii ($>$ 30 \Mpch) is simply due to excess
  halo clustering around galaxy groups and clusters.

However, the most obvious feature of this plot is the significant drop
in dust mass per galaxy at small groupcentric radii, especially for
the more massive cluster bin. The most massive clusters have dust to
galaxy ratios $\sim$ 15 $\%$ of the value far from the cluster, while
small groups have $\sim$ 80 $\%$ of the value. These numbers are not
significantly different from the dust to gas ratios we found earlier.

There are two explanations for this reduction in the dust to galaxy
ratio. First, as we have discussed, the dust could be destroyed by
thermal sputtering in hot gas.  Within a fixed physical radius, this
sputtering would be more effective in clusters than in groups, since
the average density of the hot gas is substantially higher.  Second,
it is also possible that galaxies near clusters and groups may be
fundamentally different in their dust creation properties. It is
expected that most of the dust creation occurs in supernova, and it
has been shown through extensive observations that groups and clusters
have significantly lower star formation rates than isolated
environments \citep{gomez, balogh_sdss,weinmann, pasquali} and that
difference extends to at least z=1
\citep{wilmancnoc,gerke,balogh_color}.

Unfortunately, the star formation rates in the large scale
environments around groups and clusters have not been studied as
extensively, especially at higher redshift. \citet{lewis} have
attempted to quantify this in the low redshift universe using a sample
of massive clusters. They find that the mean star formation rate of
galaxies is below the field value as far out as $\sim$ 7-10 virial
radii. Further, they see that the average star formation rate at 1
virial radius is 70 $\%$ the isolated value. However, even despite
this, the key to the amount of dust created is actually the stellar
mass, which is essentially the integrated star formation rate. Since
galaxy groups are actually quite efficient at producing stellar mass
\citep{parker, balogh_mass}, this suggests that the main driver of
lower dust to galaxy ratios near clusters is not due to reduced dust
creation. In effect, because the bulk of star formation/dust creation
occurs at higher redshift the current cluster and field star formation
rates are not as relevant.

\subsection{Impact on other science}

We have shown that there is a significant and large scale variation in
the dust reddening associated with galaxy groups and
clusters. Although studying the properties of this dust and its
implications for dust creation and destruction properties was the main
goal of the paper, in this section we briefly assess the impact this
reddening and attenuation of light might have on other studies.

\citet{menard_sn} have shown that not correcting for the impact of
dust on background ``standard candle'' supernova can bias the
measurement of $\Omega_{\rm m}$ at the few percent level. Here we
point out an additional concern in the measuring of cosmological
parameters. Our measurements imply that variations in attenuation by
dust can be as high as 0.04 magnitudes in the $V$ band (and higher in
bluer bands). The next generation of large scale photometric surveys,
such as LSST and DES, which principally are driven by the desire to
probe the cosmological parameters, have design goals of 0.01 magnitude
zeropoint stability from field to field \citep[e.g.][]{lsst,
tucker}.  Our measurements suggest this will not be
possible to achieve without accounting for the dust in and around
groups and clusters.

The dust attenuation may also be a problem for very low surface
brightness features in clusters, such as the intra-cluster and -group
medium.  \citet{zackrisson} have shown that because the dust
extinguishes background light, measuring the sky background, of which
the extragalactic background is a small part, some annulus from the
cluster would lead to an over-subtraction. Crucially, this can lead to
an underestimate of the intracluster light by as much as 1.3
magnitudes at an observed surface brightness of 30
magnitudes/arcsec$^2$ for attenuations of $A_V$=0.05. 30
magnitudes/arcsec$^2$ is close to the faint limit of the measurements
of intracluster light by \citet{zibetti}. While an attenuation of
$A_V$=0.05 is not unreasonable given our results, as we have shown,
the dust attenuation is dependent on the cluster-centric
radius. \citet{zibetti} et al. measured the sky background in a
100-kpc thick ring with an inner radius of 1 Mpc centered on their
clusters. Although this is below the spatial resolution of our
measurements, we do note that our most massive cluster varies only
from $A_V$ $\sim$ 0.042 at 1 \Mpch to $\sim$ 0.034 at 3 \Mpch. In
other words, we would expect that the variation in dust attenuations
is not large ($<$ 0.005 mag), and therefore does not induce a
significant underestimate in the intracluster light.

\section{Conclusions}\label{conclusions}
 We have used a sample of spectroscopically identified high redshift
 quasars to probe the dust content of 70,000 uniformly selected galaxy
 groups and clusters. We have used the resulting colour excesses, in
 conjugation with synthetic dust models, to estimate the excess
 surface mass density around these clusters. Finally we have compared
 the resulting dust distribution with the excess galaxies around the
 same groups and clusters. Our findings are as follows.

\begin{itemize}
\item We have shown that there exists a large scale distribution of
  dust centered on groups and clusters with masses as low as M $\sim$
  10$^{12.5}$ \Mdoth, and which extends 30 \Mpch from the group
  centre. The wavelength-dependence of this extinction is consistent
  with that expected for interstellar dust.

\item We find that the dust must be distributed relatively uniformly,
  with a covering fraction on the order of 10 $\%$ to explain the
  excess color as a function of percentile of the color distribution.

\item We find that the halo mass dependence of the dust content is
much smaller than would be expected by a simple scaling, implying that
the dust-to-gas ratio of the most massive clusters ($\sim$ 10$^{14}$
\Mdoth) is $\sim$ 3$\%$ of the local ISM value, while in small groups
($\sim$ 10$^{12.7}$ \Mdoth) it is $\sim$ 55$\%$ of the local ISM
value.

\item We find that the implied dust-to-galaxy ratio falls significantly
  closer to the group and cluster center. This reduction in the dust
  to galaxy ratio has a significant halo mass dependence, such that
  the more massive groups and clusters show a stronger reduction. This
  suggests that either dust is destroyed by thermal sputtering of the
  dust grains by the hot, dense gas or the intrinsic dust production
  is reduced in these galaxies.

\end{itemize}

\section*{Acknowledgments}

We thank the referee, Stefano Zibetti, for suggestions which improved
the paper. In particular, we thank him for clarifying the (non)-effect
of foreground cluster emission. Further, we thank he and David Wilman
for independent suggestions which led to Section 3.4. We also
acknowledge useful conversations with Brice M\'enard, David Gilbank,
Mike Hudson, David Johnston, Clif Kirkpatrick and Carolyn McCoey. MLB
acknowledges support from an NSERC Discovery Grant.

\bibliography{ms1}

\end{document}